\documentclass[a4paper, pra, twocolumn,superscriptaddress,showpacs,english]{revtex4-1}
\usepackage[T1]{fontenc}
\usepackage[latin9]{inputenc}
\usepackage{float}
\usepackage{amssymb}
\usepackage{amsmath}
\usepackage{graphicx}
\usepackage{babel}
\usepackage{color}
\def\cp#1{\mathbf{#1}}

\begin{document}

\title{Polarons and Molecules in a Fermi Gas with Orbital Feshbach Resonance}

\author{Jin-Ge Chen}
\affiliation{Department of Physics, Renmin University of China, Beijing 100872, China}
\affiliation{Beijing Key Laboratory of Opto-electronic Functional Materials and Micro-nano Devices,
Renmin University of China, Beijing 100872, China}
\author{Tian-Shu Deng}
\affiliation{Key Laboratory of Quantum Information, University of Science and Technology of China, Chinese Academy of Sciences, Hefei, Anhui, 230026, China}
\affiliation{Synergetic Innovation Center of Quantum Information and Quantum Physics, University of Science and Technology of China, Hefei, Anhui 230026, China}
\author{Wei Yi}
\email{wyiz@ustc.edu.cn}
\affiliation{Key Laboratory of Quantum Information, University of Science and Technology of China, Chinese Academy of Sciences, Hefei, Anhui, 230026, China}
\affiliation{Synergetic Innovation Center of Quantum Information and Quantum Physics, University of Science and Technology of China, Hefei, Anhui 230026, China}
\author{Wei Zhang}
\email{wzhangl@ruc.edu.cn}
\affiliation{Department of Physics, Renmin University of China, Beijing 100872, China}
\affiliation{Beijing Key Laboratory of Opto-electronic Functional Materials and Micro-nano Devices,
Renmin University of China, Beijing 100872, China}

\begin{abstract}
We study the impurity problem in a gas of $^{173}$Yb atoms near the recently discovered orbital Feshbach resonance. In an orbital Feshbach resonance, atoms in the electronic ground state $^1S_0$ interact with those in the long-lived excited $^3P_0$ state with magnetically tunable interactions. We consider an impurity atom with a given hypferine spin in the $^3P_0$ state interacting with a single-component Fermi sea of atoms in the ground $^1S_0$ manifold. Close to the orbital Feshbach resonance, the impurity can induce collective partilce-hole excitations out of the Fermi sea, which can be regarded as the polaron state. While as the magnetic field decreases, a molecular state becomes the ground state of the system. We show that a polaron to molecule transition exists in $^{173}$Yb atoms close to the orbital Feshbach resonance. Furthermore, due to the spin-exchange nature of the orbital Feshbach resonance, the formation of both the polaron and the molecule involve spin-flipping processes with interesting density distributions among the relevant hyperfine spin states. We show that the polaron to molecule transition can be detected using Raman spectroscopy.
\end{abstract}
\pacs{67.85.Lm, 03.75.Ss, 05.30.Fk}

\maketitle

\section{Introduction}
In recent years, alkaline-earth and alkaline-earth-like atoms have attracted much research interest. With two valence electrons, these atoms acquire interesting features such as long-lived electronically excited states and the separation of nuclear- and electronic-spin degrees of freedom in the so-called clock states. These features have been extensively investigated for proposals and applications in precision measurements~\cite{AE1,AE2,AE3}, as well as quantum information and quantum simulation~\cite{congjun03,AE5,AE4,ofr1,ofr2,ofr3}. However, it has long been considered difficult to realize a stable, strongly-interacting gas of alkaline-earth or alkaline-earth-like atoms, due to the lack of ground-state magnetic Feshbach resonances. Although the interactions can be tuned via optical Feshbach resonances, it is only at the cost of severe atom losses~\cite{opticalFR1,opticalFR2,opticalFR3}.

In this context, the recently discovered orbital Feshbach resonance (OFR) in $^{173}$Yb atoms has enabled the study of strongly-interacting systems using alkaline-earth and alkaline-earth-like atoms, and thus greatly enriches the scope of quantum simulation in these systems~\cite{ren1,ofrexp1,ofrexp2}. Like the magnetic Feshbach resonances in alkaline atoms~\cite{chinreview}, the OFR derives from the resonant scattering between an open channel and a closed channel. In $^{173}$Yb atoms for example, the open channel corresponds to two atoms occupying the $|g\downarrow\rangle$ and the $|e\uparrow\rangle$ states respectively, where $|g\rangle$ ($|e\rangle$) corresponds to the $^1S_0$ ($^3P_0$) state, and $|\downarrow\rangle$ and $|\uparrow\rangle$ represent two different nuclear spin states in the hyperfine manifold of the clock states $^1S_0$ and $^3P_0$. On the other hand, the closed channel corresponds to two atoms occupying the $|g\uparrow\rangle$ and the $|e\downarrow\rangle$ states respectively. Note that since $J=0$ for $|g\rangle$ and $|e\rangle$, the nuclear spins and the electronic spins are decoupled. As a result, the two-body interaction at short ranges can occur either in the electronic spin-singlet channel $|-\rangle \equiv \frac{1}{2}(|ge\rangle-|eg\rangle)\otimes(|\downarrow\uparrow\rangle+|\uparrow\downarrow\rangle)$, or in the electronic spin-triplet channel $|+ \rangle \equiv \frac{1}{2}(|ge\rangle+|eg\rangle)\otimes(|\downarrow\uparrow\rangle-|\uparrow\downarrow\rangle)$~\cite{ofr1,ofr2,ofr3}. The short-range interactions are thus off-diagonal in the basis of open and closed channels, and couple the two channels in the form of an inter-orbital nuclear-spin-exchange interaction. Furthermore, in the presence of a finite magnetic field, the differential Zeeman shift between $|g\rangle$ and $|e\rangle$ should give rise to a tunable energy difference between the open and the closed channels~\cite{zeemanshift1,zeemanshift2,ren1}. Thus, a resonant scattering can occur when a shallow bound state in the open channel is tuned to the two-body scattering threshold. The OFR is first proposed theoretically in Ref.~\cite{ren1}, and subsequently confirmed experimentally using $^{173}$Yb atoms. Further studies show that the OFR is a narrow resonance in terms of the resonance width, but a wide one in terms of the magnetic field~\cite{ren2}. The latter makes the OFR easily accessible in $^{173}$Yb atoms.

A particularly interesting feature of OFR is the spin-exchange nature of the two-body interaction potential. While this interesting feature has seen applications in quantum information, it would be interesting to see its effects in a many-body setting. A promising scenario to examine is the impurity problem close to an OFR. In alkaline atoms, impurity problems across the conventional magnetic Feshbach resonance have been extensively studied in recent years~\cite{mitpolaron,enspolaron,grimmpolaron,kohlpolaron,impurityreview1,impurityreview2,mag3,mag2,parishprl,wypolaron1,wypolaron2,3body5,3body6}. These theoretical and experimental investigations provide physical insights into the underlying many-body system in the highly-polarized limit, and serve to bridge few- and many-body physics. In systems with OFR, it is natural to expect that similar studies should offer valuable information of the system on both the few- and the many-body levels.

In this work, we consider an impurity atom in an excited state $|e\uparrow\rangle$ interacting with a Fermi sea of atoms in $|g\downarrow\rangle$ of the ground-state manifold. In the presence of a finite magnetic field, the OFR mechanism leads to a tunable interaction strength between the impurity atom and atoms in the Fermi sea. As a result, the impurity atom can either induce collective particle-hole excitations out of the Fermi sea, or form tightly bound molecular states, depending on the parameters. Typically, this should lead to the so-called polaron to molecule transition as the interaction strength is tuned. Due to the inter-orbital spin-exchange interactions of OFR, the formations of both the polaron and the molecular states feature novel spin-flipping processes. For example, an atom in the $|g\downarrow\rangle$ state and another in the $|e\uparrow\rangle$ state are scattered to the $|g\uparrow\rangle$ and the $|e\downarrow\rangle$ states. As a result, for the polaron, besides the conventional spin-conserving particle-hole excitations, one should also have spin-flipping excitations, under which previously empty hyperfine states become occupied. While in the molecular state, two-body bound state also emerges in the originally unoccupied states. Adopting the Chevy-like ansatz, we demonstrate that these spin-flipping processes can actually be quite important and lead to considerable weight in the corresponding variational wave functions. We then show that a polaron to molecule transition exists in the system close to the OFR, whose exact location depends on the atomic density of the system. We propose to detect the polaron to molecule transition using Raman spectroscopy, by coupling population in the originally empty hyperfine states to a by-stander state. The polaron to molecule transition gives rise to a sharp signature in the ground-state Raman spectra of the system. Our results can be checked under current experimental conditions.

The remainder of this paper is organized as follows. In Sec.~\ref{sec:H}, we present the formalism of the impurity problem under consideration. The solutions of the molecule and the polaron states throughout the resonance region are discussed in Sec.~\ref{sec:mol} and Sec.~\ref{sec:pol}, respectively. With these knowledge, we study the polaron--molecule transition, and propose a detection scheme using Raman spectroscopy in Sec.~\ref{sec:transition}. Finally, we summarize the main findings in Sec.~\ref{sec:con}.

\section{Impurity problem in Fermi gases with orbital Feshbach resonance}
\label{sec:H}
\begin{figure}[t]
\centering{}
\includegraphics[width=0.7\columnwidth]{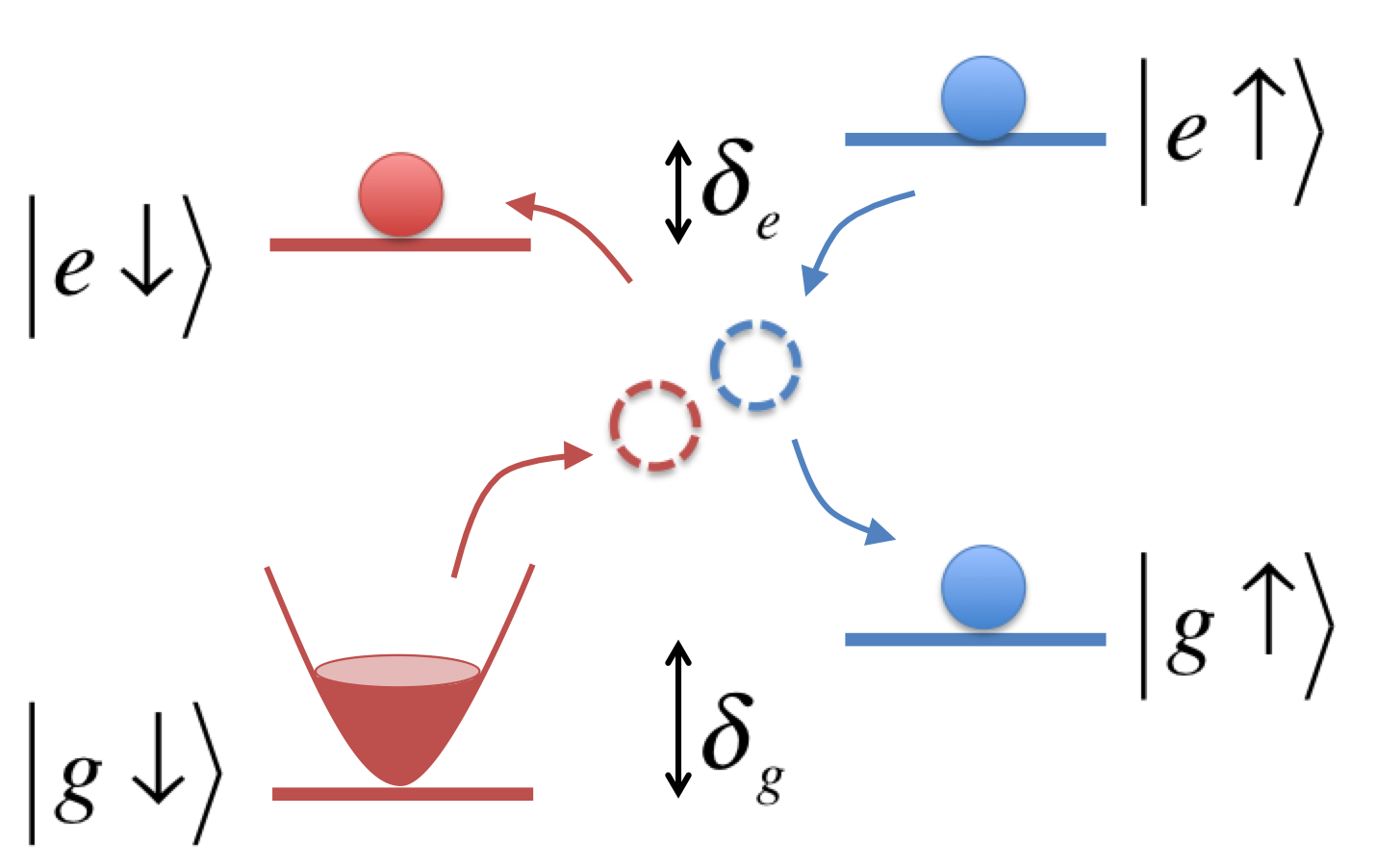}
\caption{(Color online) Level diagram of orbital Feshbach resonance in alkaline-earth-like
atoms. An impurity of $| e \uparrow \rangle$ is immersed in a majority Fermi sea of
$| g \downarrow \rangle$ atoms, and can be scattered to the other two atomic states forming
the closed channel via interaction. $\delta_g$ and $\delta_e$ are the Zeeman shifts of the $|g\rangle$ and $|e\rangle$ manifolds, respectively.}
\label{fig:scheme}
\end{figure}

We consider a Fermi gas of alkaline-earth (-like) atoms across its OFR,
where the two nuclear spin states and the two orbital states combined together to form the
open and the closed channels with a magnetically tunable relative energy.
The Hamiltonian of the system can be written as~\cite{ren1,ren2}
\begin{eqnarray}
\label{eqn:H}
H &=& \sum_{\bf k} \left(\epsilon_{\bf k} + \frac{\delta}{2}\right)
\left( a_{{\bf k} g \downarrow}^\dagger a_{{\bf k} g \downarrow}
+ a_{{\bf k} e \uparrow}^\dagger a_{{\bf k} e \uparrow} \right)
\nonumber \\
&&+
\sum_{\bf k} \epsilon_{\bf k} \left( a_{{\bf k} g \uparrow}^\dagger a_{{\bf k} g \uparrow}
+ a_{{\bf k} e \downarrow}^\dagger a_{{\bf k} e \downarrow} \right)
+ H_{\rm int},
\end{eqnarray}
where $a_{{\bf k} p \sigma}^\dagger$ and $a_{{\bf k} p \sigma}$ ($p=e,g$ and $\sigma = \uparrow, \downarrow$) are fermionic operators associated with the corresponding states with three-dimensional linear momentum ${\bf k}$.
Due to the difference of Land\'e factors in the $|g\rangle$ and $|e \rangle $ orbitals, the Zeeman shifts $\delta_g$ and $\delta_e$ between two hyperfine states within each electronic manifold are distinct, as illustrated in Fig.~\ref{fig:scheme}. As a consequence, the differential Zeeman shift $\delta = \delta_e - \delta_g$ between the open channel, consisting of the $| g \downarrow \rangle$ and $| e \uparrow \rangle$ states, and the closed channel composed with the $| g \uparrow \rangle$ and $| e \downarrow \rangle$ states, can be tuned by sweeping the magnetic field. An OFR occurs as one of the closed-channel bound states moves across the open-channel threshold, or vice versa~\cite{ren1, ofrexp1, ofrexp2}. Notice that by writing down Eq. (\ref{eqn:H}), we shift the single-particle dispersions such that they can be written in a symmetric form, with $|g \uparrow \rangle$ and $| e \downarrow \rangle$ at the zero-energy reference, while with $|g \downarrow \rangle$ and $| e \uparrow \rangle$ both detuned by $\delta/2$.

By using the basis of electronic singlet and triplet channels $|\pm\rangle$,
the interaction term takes the following form
\begin{eqnarray}
\label{eqn:Hint}
H_{\rm int} &=& \sum_{\bf q} \left(\frac{g_+}{2} \hat{A}_{{\bf q}, +}^\dagger \hat{A}_{{\bf q}, +}
+ \frac{g_-}{2} \hat{A}_{{\bf q}, -}^\dagger \hat{A}_{{\bf q}, -} \right)
\end{eqnarray}
with the operators defined as
\begin{eqnarray}
\label{eqn:A}
\hat{A}_{{\bf q}, +} &=& \sum_{\bf k} \left(a_{-{\bf k}+ {\bf q}, g \downarrow} a_{{\bf k}+ {\bf q}, e \uparrow}  -
a_{-{\bf k}+ {\bf q}, g \uparrow} a_{{\bf k}+ {\bf q}, e \downarrow} \right),
\nonumber \\
\hat{A}_{{\bf q}, -} &=& \sum_{\bf k} \left(a_{-{\bf k}+ {\bf q}, g \downarrow} a_{{\bf k}+ {\bf q}, e \uparrow}  +
a_{-{\bf k}+ {\bf q}, g \uparrow} a_{{\bf k}+ {\bf q}, e \downarrow} \right).
\end{eqnarray}
The interaction strength $g_\pm$ are related to the physical ones via the standard renormalization relation
$1/g_\pm = 1/{g_\pm^p} - \sum_{\bf k} 1/(2 \epsilon_{\bf k})$, where $g_\pm^p = 4 \pi \hbar^2 a_\pm/m$ 
with $a_\pm$ the corresponding $s$-wave scattering lengths and $m$ the atomic mass.

The fermion impurity problem, by definition, is to impose a minority impurity atoms against a majority Fermi sea formed
by fermions of another type. In this paper, without loss of generality, we assume that the system composes of a Fermi sea of
$N$ non-interacting $|g \downarrow \rangle$ particles and an impurity $|e \uparrow \rangle$ particle,
as schematically illustrated in Fig.~\ref{fig:scheme}. Notice that the other configuration, with
$|g \uparrow \rangle$ as the majority fermion and $| e\downarrow \rangle$ as the impurity, can be mapped
to the present scenario by simply changing the sign of $\delta$ and reversing the definition of the open and the closed channels.

\section{The Molecular state}
\label{sec:mol}
\begin{figure}[t]
\centering{}
\includegraphics[width=0.98\columnwidth]{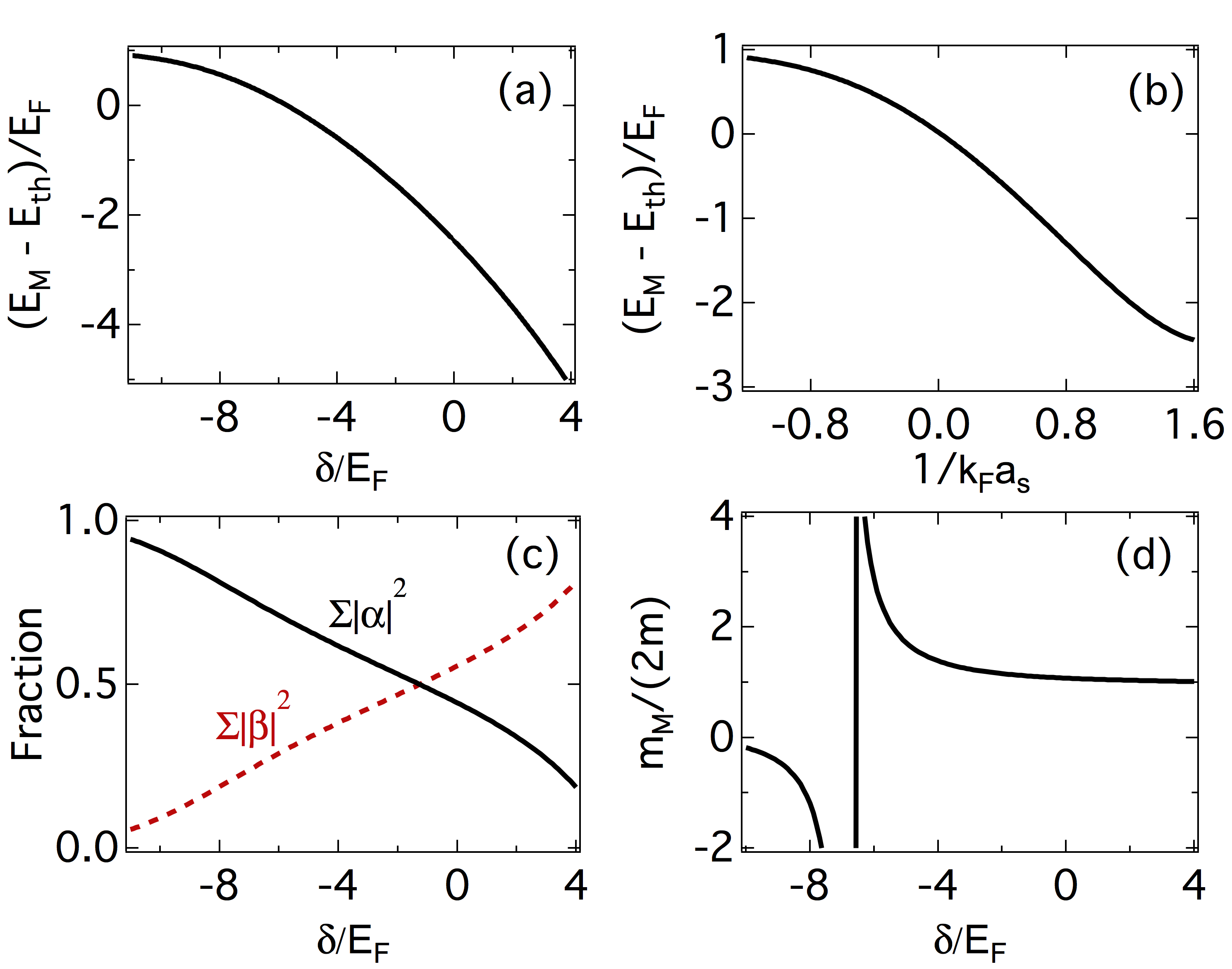}
\caption{(Color online) (a-b) The eigen energy shifted by the threshold energy $E_{\rm th}$,
(c) the distribution of wave functions in the open and the closed channels, and (d) the effective
mass are plotted for the shallow molecular state. Parameters used here are compatible with
a gas of ${}^{173}$Yb atoms with a number density of $n = 2 \times 10^{13} {\rm cm}^{-3}$.
The resonance takes place at $\delta/E_F \approx -5.8$ and the position $\delta/E_F = 0$ corresponds 
to $1/(k_F a_s) \approx 1.67$ within this parameter set. Notice that the effective mass diverges 
and changes sign at $\delta/E_F \approx -6.6$, which resides on the BCS side of the resonance 
with $1/(k_F a_s) \approx -0.18$.}
\label{fig:mol}
\end{figure}

In a molecular state, one of the majority fermions is pulled out of the Fermi sea and forms a bound state
with the impurity fermion, leaving the rest $N-1$ majority particles in the Fermi sea. A trial
wave function can be written as
\begin{eqnarray}
\label{eqn:molwf}
|M\rangle_{\bf Q} &=& \sum_{|{\bf k}| > k_F} \alpha_{\bf k}
a_{{\bf Q}-{\bf k},e\uparrow}^\dagger a_{{\bf k},g\downarrow}^\dagger | g\downarrow \rangle_{N-1}
\nonumber \\
&& +
\sum_{\cp k} \beta_{\bf k}
a_{{\bf Q}-{\bf k},e\downarrow}^\dagger a_{{\bf k},g\uparrow}^\dagger | g\downarrow \rangle_{N-1},
\end{eqnarray}
where ${\bf Q}$ denotes the center-of-mass momentum, and $| g \downarrow \rangle_{N-1}$ represents
the Fermi sea of $N-1$ majority particles. Notice that in the expression above, the two particles
constituting the molecular state can be either in the open or in the closed channel, with the corresponding
coefficients $\alpha_{\bf k}$ or $\beta_{\bf k}$. This ansatz has been employed in the discussion of fermion impurity problems in different dimensions close to a magnetic Feshbach resonance, and is usually referred to as the
bare molecular state, as all particle-hole fluctuations atop the Fermi sea are neglected~\cite{impurityreview1,impurityreview2}.

The Schr\"odinger equation for the ansatz wave function Eq. (\ref{eqn:molwf}) is then given by
\begin{eqnarray}
\label{eqn:molsch}
H |M\rangle_{\bf Q} = {\tilde E}_M({\bf Q}) | M \rangle_{\bf Q},
\end{eqnarray}
with ${\tilde E}_M({\bf Q})$ the eigen energy. By substituting the ansatz wave function into the
Hamiltonian Eq. (\ref{eqn:H}) and matching terms, we can obtain a set of linear equations for
the coefficients $\alpha_{\bf k}$ and $\beta_{\bf k}$, after dropping some higher-order terms. This can also be done by evaluating the
expectation value of ${}_{\bf Q} \langle M| H - {\tilde E}_M({\bf Q}) |M\rangle_{\bf Q}$, and taking derivatives
with respect to $\alpha_{\bf k}$ and $\beta_{\bf k}$. The coefficient equations can be grouped into the following two equations
\begin{eqnarray}
\label{eqn:molcoeff}
&&\left(1 + \frac{g_- + g_+}{2} \Theta_{\bf Q} \right) \sum_{|{\bf k}_1| > k_F} \alpha_{{\bf k}_1}
+
\frac{g_- - g_+}{2} \Theta_{\bf Q} \sum_{{\bf k}_1} \beta_{{\bf k}_1} =0,
\nonumber \\
&&\frac{g_- - g_+}{2} \Theta_{\bf Q}^\prime \sum_{|{\bf k}_1| > k_F} \alpha_{{\bf k}_1}
+ \left( 1 + \frac{g_- + g_+}{2}\Theta_{\bf Q}^\prime \right) \sum_{{\bf k}_1} \beta_{{\bf k}_1} =0,
\nonumber \\
\end{eqnarray}
where the parameters are defined as
\begin{eqnarray}
\Theta_{\bf Q} &=& \sum_{|{\bf k}|>k_F} \frac{1}{\epsilon_{\bf k} + \epsilon_{{\bf Q}-{\bf k}} + \delta - E_M},
\nonumber \\
\Theta_{\bf Q}^\prime &=& \sum_{\bf k} \frac{1}{\epsilon_{\bf k} + \epsilon_{{\bf Q}-{\bf k}} -E_M}.
\end{eqnarray}
To simplify notation, we shift the energy reference
$E_M = {\tilde E}_M - \sum_{|{\bf k}|<k_F} (\epsilon_{\bf k} + \delta/2)$.
Note that in this reference, the threshold energy $E_{\rm th} = E_F + \delta$ for a non-interacting impurity
$|e\uparrow \rangle$ immersed in a majority $|g\downarrow \rangle$ Fermi sea with $N$ particles.
The secular equation obtained from Eq. (\ref{eqn:molcoeff}) hence leads to a closed equation
for the eigen energy of the molecular state, which takes the following form after renormalization
\begin{eqnarray}
\label{eqn:moleq}
&&\frac{1}{g_-^p g_+^p} + \frac{1}{2}\left( \frac{1}{g_-^p} +  \frac{1}{g_+^p} \right)
\left( \Theta_{\bf Q} + \Theta_{\bf Q}^\prime - 2 \Lambda_c \right)
\nonumber \\
&& \hspace{3cm}
+ (\Theta_{\bf Q} - \Lambda_c)(\Theta_{\bf Q}^\prime -\Lambda_c)
= 0.
\end{eqnarray}
with $\Lambda_c \equiv \sum_{\bf k} 1/(2\epsilon_{\bf k})$.

We first discuss the molecular state with a zero center-of-mass momentum ${\bf Q} = 0$. By numerically solving
Eq. (\ref{eqn:moleq}), we find two solutions for $E_M$: one shallow branch with energy close to
the threshold energy $|E_M - E_{\rm th}| \sim E_F$, and another deeply bound state with energy
$|E_M - E_{\rm th}| \gg E_F$.
Figures~\ref{fig:mol}(a) and \ref{fig:mol}(b) show the variations of the
shallow molecular energy shifted by the threshold energy $E_M - E_{\rm th}$ as functions of $\delta$
and $1/(k_F a_s)$, respectively, where the $s$-wave scattering length is given by the relation~\cite{ren2}
\begin{eqnarray}
\label{eq:as}
a_s = \frac{-a_{s0} + \sqrt{m |\delta| / \hbar^2}(a_{s0}^2 - a_{s1}^2)}{a_{s0} \sqrt{m|\delta|/\hbar^2} -1}
\end{eqnarray}
with $a_{s0} = (a_+ + a_-)/2$ and $a_{s1} = (a_- - a_+)/2$. Throughout this paper, as a typical example,
we consider a gas of ${}^{173}$Yb atoms with a number density of $n = 2 \times 10^{13} {\rm cm}^{-3}$
unless otherwise specified, and use $E_F$ as the energy unit with $m=1/2$, $\hbar = 1$ and $k_F =1$.
The scattering lengths are taken as $a_+ = 1900 a_0$ and $a_- = 200 a_0$ with $a_0$ the Bohr
radius~\cite{ofrexp2,ofrexp2,ren2}. With these parameters, the energy of the deep bound state almost 
remains as a constant $~160 E_F$, which is far detuned from the threshold. This is consistent with 
the physical origins of the deep and the shallow states. In fact, as the shallow state is intimately related 
to the OFR tuned by $\delta$, the deep molecular state is supported by the positive background interaction 
in the $|-\rangle$ channel, and is therefore far detuned from the threshold.
\begin{figure}[t]
\centering{}
\includegraphics[width=0.98\columnwidth]{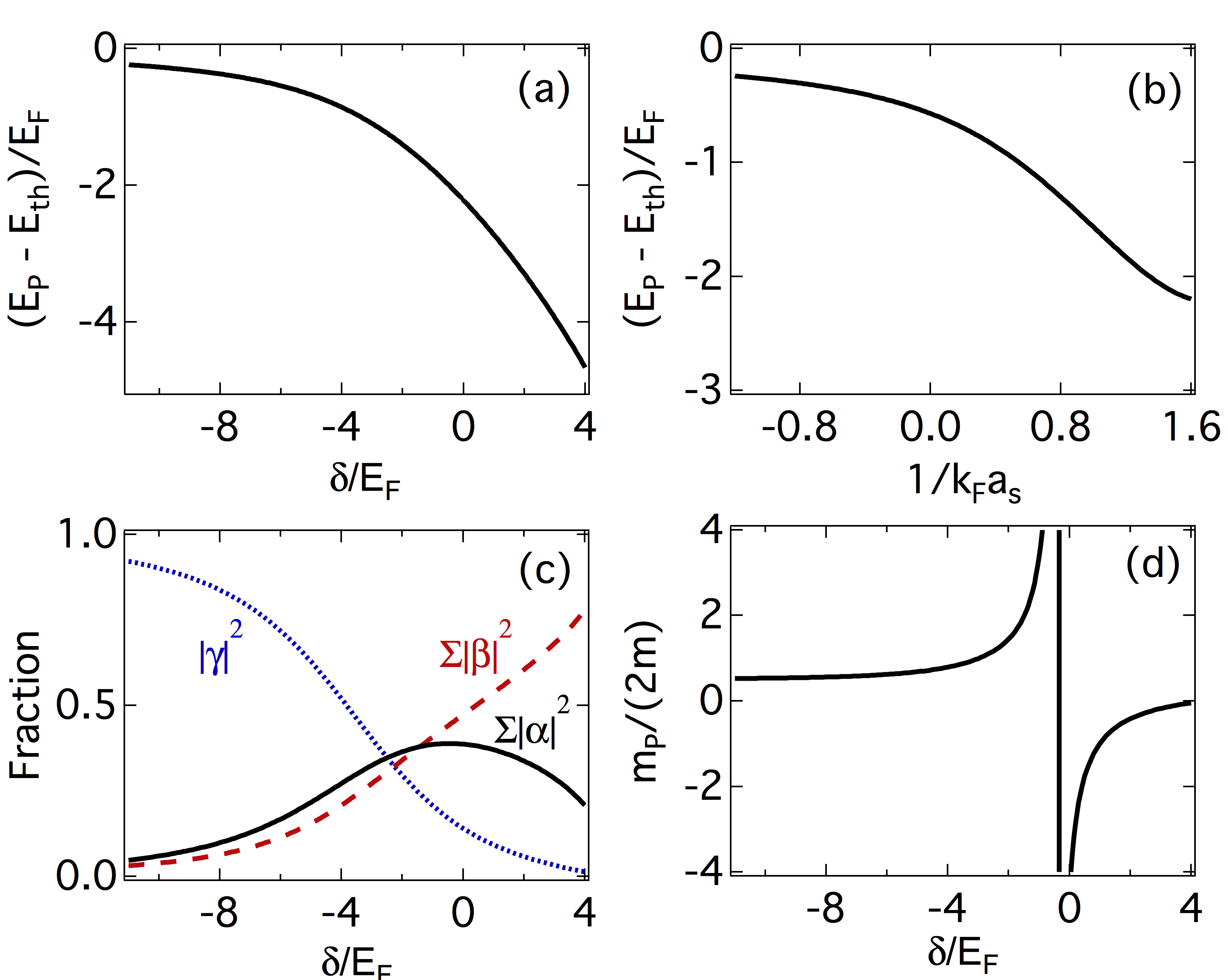}
\caption{(Color online) (a-b) Energy shifted from the threshold energy,
(c) Fraction of wave functions within the open-, closed-, and bare-particle-channels,
and (d) the effective mass for the shallow polaron state with a fixed momentum $Q=0$.
The effective mass diverges at $\delta / E_F \approx -0.35$, corresponds to $1/(k_F a_s) \approx 1.40$.
Parameters used are the same as in Fig.~\ref{fig:mol}.}
\label{fig:pol}
\end{figure}

By substituting the solutions of eigen energy into the coefficient equations, we can solve for the
coefficients within the ansatz wave function Eq. (\ref{eqn:molwf}). In Figs.~\ref{fig:mol}(c),
we show the fraction of population in the open and the closed channels for the shallow state.
While the closed-channel fraction $\sum_{\bf k} |\beta_{\bf k}|^2$ increases with $\delta$ and
becomes dominant for large $\delta$, the open-channel fraction $\sum_{\bf k} |\alpha_{\bf k}|^2$
follows the opposite trend. This observation can be understood by noticing that at a large positive $\delta$,
the open channel is largely detuned above the closed channel as shown in Fig~\ref{fig:scheme},
leading to a negligible fraction in the open channel. As a comparison, the deeply bound state is almost
equally populated between the open and the closed channels, reflecting the fact that the $|-\rangle$ state
is a linear combination of the open and the closed channels with equal weights.

We then turn to the general case of ${\bf Q} \neq 0$. By considering a small deviation of $\cp Q$ away
from zero, we can calculate the effective masses of the molecules. As shown in Fig.~\ref{fig:mol}(d),
the effective mass for the shallow-branch molecule increases from a limiting value of unity
in the deep BEC limit, where the molecule is essentially a structureless boson with unit mass,
diverges at some finite detuning, and becomes negative when moving further toward the BCS side.
This trend of variation is qualitatively consistent with previous studies of molecular state for
both wide~\cite{leyronas-09} and narrow~\cite{massignan-12, castin-12} magnetic Feshbach resonances.
Quantitatively, the effective mass diverges at the point $\delta/E_F \approx -6.6$, which resides
on the BCS side of resonance with $1/(k_F a_s) \approx -0.18$. 
While the divergence of the effective mass here suggests
the acquisition of a finite center-of-mass momentum for the molecules, this result also implies that
the OFR is a narrow resonance with a finite effective range
potential, as the divergence of molecular effective mass typically occurs
on the BEC side of a wide magnetic Feshbach resonance~\cite{massignan-12, castin-12}.  
As a comparison, the molecular state in the deep
branch remains to be a structureless boson with $m_M/(2m) \approx 1$ across the entire resonance regime.

\section{The Polaron state}
\label{sec:pol}

\begin{figure}[t]
\centering{}
\includegraphics[width=0.98\columnwidth]{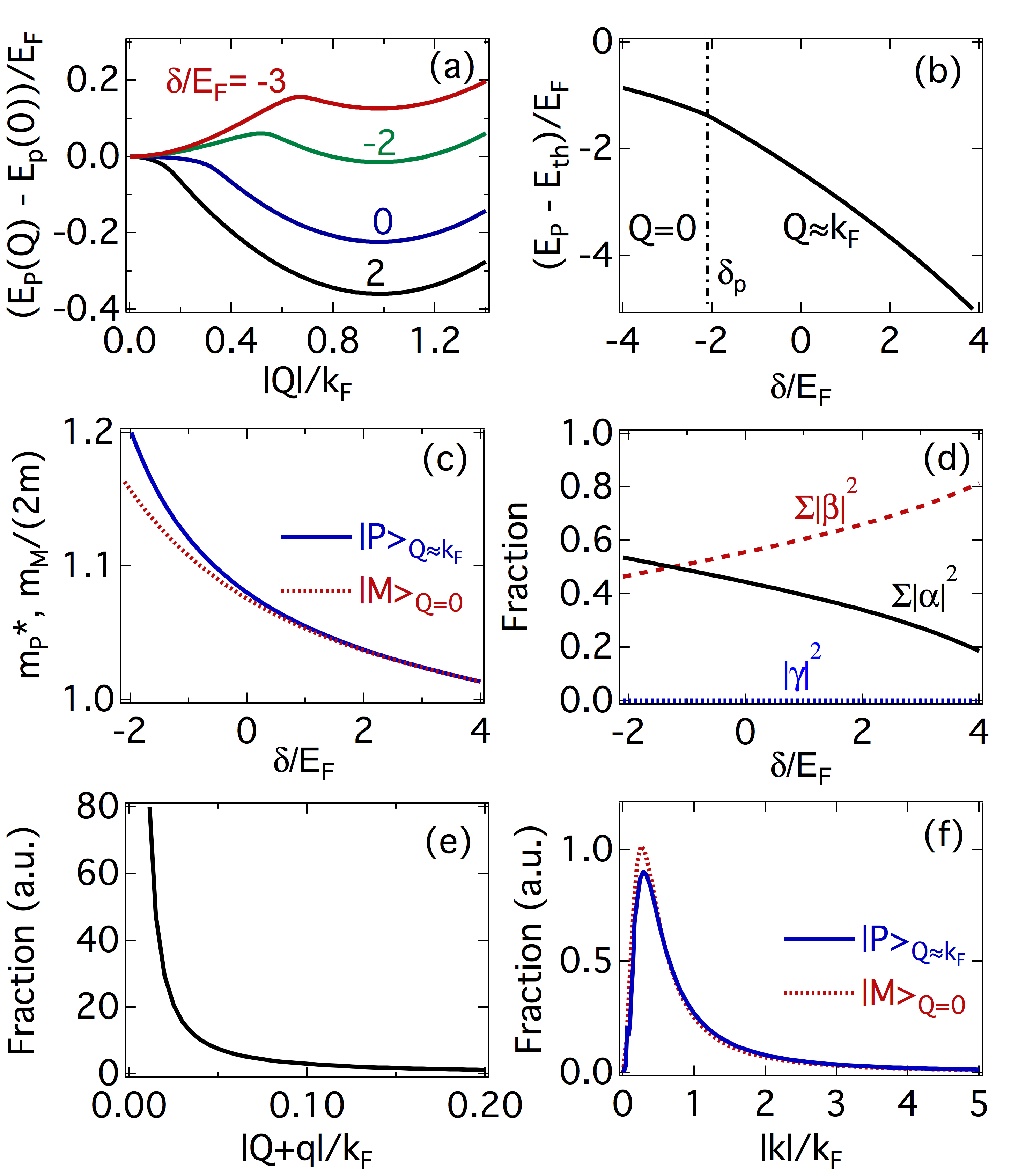}
\caption{(Color online) (a) Dispersion of the shallow polaron state for various values
of $\delta$. The zero-momentum $| P \rangle_{Q=0}$ state is the ground state for large negative
$\delta$, which becomes a metastable state by passing through a first-order-like transition
around $\delta_p/E_F \lesssim -2.1$. It eventually becomes unstable by further
increasing $\delta$. All dispersion curves are shifted with respect to their corresponding
zero-momentum state energy, so that they can be compared in the same plot.
(b) The transition between two locally stable polaron states. The polaron state
with $|{\bf Q}| \sim k_F$ resembles the zero-center-of-mass-momentum molecular state
with increasing $\delta$, as can be seen from (c) the ``effective mass'' obtained by fitting the
dispersion around the global minimum, and (d) the fraction of wavefunction distribution in the
open-, closed-, and bare-particle-channels.
In the BEC limit of $\delta/E_F = 4$, (e) the closed-channel fraction as a function
of $|{\bf Q} + {\bf q}|$ suggests that the major contribution to the hole is close to the
Fermi surface, while the results of momentum distribution of the $| e \downarrow \rangle$ state (f)
explicitly illustrates the similarity between the finite-momentum polaron and the
molecular state. Parameters used are the same as in Fig.~\ref{fig:mol}. }
\label{fig:polsQ}
\end{figure}

To study the polaron state of the system, we consider the following ansatz wave function
\begin{eqnarray}
\label{eqn:polwf}
|P\rangle_{\bf Q} &=& \gamma a^\dagger_{{\bf Q}e\uparrow}|g_{\downarrow}\rangle_N +
\sum_{\substack{{|{\bf k}|>k_F}\\{|{\bf q}|<k_F}}} \alpha_{{\bf k q}}
a^\dagger_{{\bf Q}+{\bf q}-{\bf k}, e\uparrow}a^\dagger_{{\bf k}g\downarrow}a_{{\bf q}g\downarrow}|g_{\downarrow}\rangle_N
\nonumber \\
&+&
\sum_{\substack{{\bf k}\\{|{\bf q}|<k_F}}} \beta_{\bf k q}
a^\dagger_{{\bf Q}+{\bf q}-{\bf k}, e\downarrow}a^\dagger_{{\bf k}g\uparrow}a_{{\bf q}g\downarrow}|g_\downarrow\rangle_N.
\end{eqnarray}
In this expression, the first term corresponds to a bare impurity and an unperturbed
Fermi sea, the second term represents a state with one pair of particle-hole excitation atop the Fermi sea in the open channel,
and the third term corresponds to the state where the majority fermion created above the Fermi surface interacts
with the impurity and both are scattered into the closed channel.

By writing down the Schr\"odinger equation $H |P\rangle_{\bf Q} = {\tilde E}_P({\bf Q}) |P\rangle_{\bf Q}$, and
following the same procedure as outlined in the previous section, we obtain
the closed equation for the eigen energy of the polaron
\begin{align}
\label{eqn:poleq}
E_P - \frac{\delta}{2} - \epsilon_{\cp Q} &= \sum_{|\cp q|<k_F}
\Bigg\{ \frac{1}{2}\left(\frac{1}{g_+^p} + \frac{1}{g_-^p}\right) +
\Gamma_{\bf Qq}^\prime - \Lambda_c
\nonumber \\
& \hspace{-1.8cm}
- \frac{1}{4}\left( \frac{1}{g_+^p} - \frac{1}{g_-^p}\right)^2
\left[{\frac{1}{2}\left(\frac{1}{g^p_+}+\frac{1}{g^p_-}\right)+ \Gamma_{\bf Qq}-\Lambda_c }\right]^{-1}
\Bigg\}^{-1},
\end{align}
where
\begin{eqnarray}
\label{eqn:polpara}
\Gamma_{\bf Qq}^\prime &=& \sum_{|\cp k|>k_F}\frac{1}{\epsilon_{\cp k}-\epsilon_{\cp q}+\epsilon_{\cp Q+\cp q-\cp k} +\frac{\delta}{2} - E_P},
\nonumber\\
\Gamma_{\bf Qq}&=& \sum_{\cp k}\frac{1}{\epsilon_{\cp k}-\epsilon_{\cp q}+\epsilon_{\cp Q+\cp q-\cp k}-\frac{\delta}{2}-E_P}.
\end{eqnarray}
Notice that the polaron energy in the expression above is also shifted by the same zero-point energy
$E_P = {\tilde E}_P - \sum_{|{\bf k}|<k_F} (\epsilon_{\bf k} + \delta/2)$.

We then analyze the polaron state by solving Eq. (\ref{eqn:poleq}) and the corresponding
coefficient equations. Similar to the molecular state, the polaron state also has two solutions with a
shallow branch close to the threshold and a deep branch with energy $|E_P - E_{\rm th}| \gg E_F$.
As the deep branch is a direct consequence of positive background scattering length in the $|-\rangle$
channel, and is not easily accessible experimentally, in the following
discussion we focus on the shallow branch. In Fig.~\ref{fig:pol}, we show the
energy, the wave function fractions of the open-, closed-, and bare-particle-channels,
as well as the effective mass for the shallow state with a
fixed momentum ${\bf Q}=0$. A key finding is that the wave function is dominated by the bare impurity sector $\gamma^2$ in the BCS limit with a large negative $\delta$, while the closed-channel contribution
$\sum_{{\bf k},|{\bf q}|<k_F} |\beta_{\bf kq}|^2$ prevails in the BEC limit. The open-channel fraction
$\sum_{|{\bf k}|>k_F,|{\bf q}|<k_F} |\alpha_{\bf kq}|^2$ remains comparable to the closed-channel fraction
around resonance, and starts to drop with increasing $\delta$ as it becomes energetically less favorable.

Next, we discuss the general situation of ${\bf Q} \neq 0$. Similar to the molecular case, we calculate effective
mass of the shallow polaron state by considering a small deviation from ${\bf Q} = 0$. From Fig.~\ref{fig:pol}(d), 
We find that the polaron effective mass approaches a limiting value of 1/2 in the BCS limit, where the 
system reduces to a non-interacting impurity atom of mass 1/2 atop a unperturbed Fermi sea. By moving 
toward the resonance point, the effective 
mass increases and presents a diverging behavior as in the molecular case. The diverging point locates at 
$\delta/E_F \approx -0.35$, which is on the BEC side of the resonance with $1/(k_F a_s) \approx 1.40$. 
To get further insight of this divergence, we extend our calculation to large $|{\bf Q}|$, and find that the 
zero-momentum polaron is stable for large negative $\delta$, becomes metastable, and eventually unstable 
with increasing $\delta$. As one can see clearly in Fig.~\ref{fig:polsQ}(a),
the dispersion of the shallow polaron acquires another local minimum around
$|{\bf Q}|/k_F \approx 1$. The competition between the two local minima hence leads to a
first-order-like transition around $\delta_p/E_F \approx -2.1$ for the parameters discussed here,
as indicated in Fig.~\ref{fig:polsQ}(b).

The emergence of a finite-momentum polaron with $|{\bf Q}|/k_F \approx 1$ can be understood by noticing that such a solution resembles the zero-momentum molecular state in the BEC limit with a large positive $\delta$. This can be confirmed by comparing the energy [Fig.~\ref{fig:transition}(a)], the ``effective mass'' [Fig.~\ref{fig:polsQ}(c)], and the wavefunction fractions [Figs.~\ref{fig:polsQ}(d-f)] between the $|{\bf Q}|/k_F \approx 1$ polaron and the shallow-branch molecule. For the $| M\rangle_{Q=0}$ state, as shown in Eq. (\ref{eqn:molwf}), the main contribution to the wave function thus corresponds to a Fermi sea of $(N-1)$ $|g\downarrow\rangle$ atoms plus
a zero center-of-mass molecule consisting of two particles in the $|g\uparrow\rangle$ and
$|e\downarrow\rangle$ states, respectively. Meanwhile, for the $| P \rangle_{|{\bf Q}| \approx k_F}$ state,
the wave function is dominated by terms with $|{\bf Q}+{\bf q}|/k_F \sim 0$, as illustrated in
Fig.~\ref{fig:polsQ}(e). Physically, this result shows that the $| P \rangle_{|{\bf Q}| \approx k_F}$
state is essentially composed of a Fermi sea of $N$ $|g\downarrow\rangle$ atoms,
with a hole very close to the Fermi surface with $|{\bf q}| \sim k_F$, and a zero center-of-mass
molecule within the closed channel. This state thus resembles the $| M\rangle_{Q=0}$ state
and becomes energetically favorable compared to the $Q=0$ polaron in the BEC limit. However, as we will show later, the $|{\bf Q}|/k_F \approx 1$ polaron is metastable against the shallow-branch molecular state.

\section{Polaron--Molecule transition}
\label{sec:transition}
\begin{figure}[t]
\centering{}
\includegraphics[width=0.98\columnwidth]{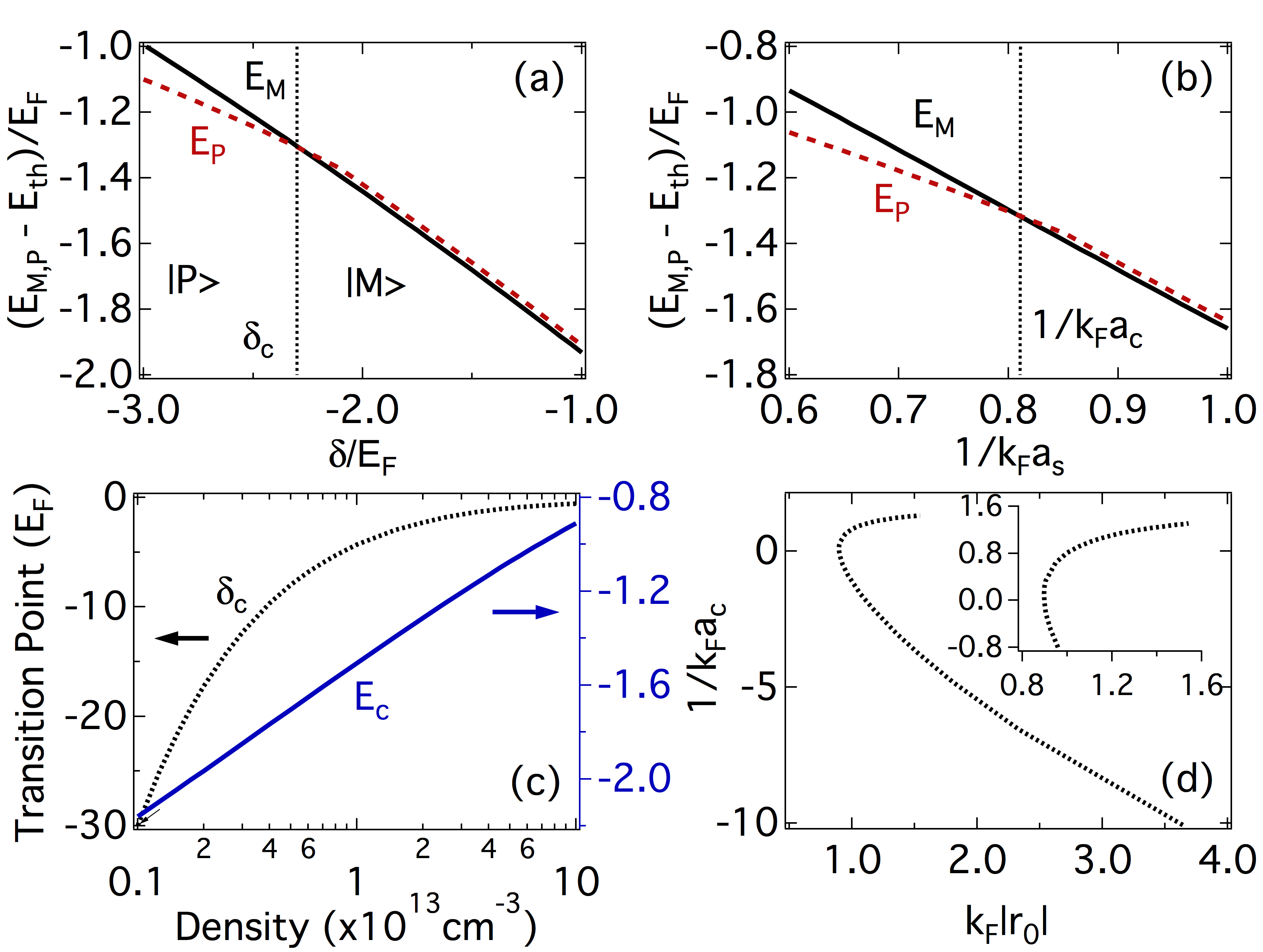}
\caption{(Color online) (a-b) Energies of the molecule (black solid) and the polaron (red dashed)
in the shallow branch. A transition in the ground states occurs at around
$\delta_c/E_F \approx -2.3$ or $1/(k_F a_c) \approx 0.81$. Note that this polaron--molecule
transition happens before the transition between the zero-momentum and finite-momentum polaron,
so that the finite-momentum polaron state is always metastable. The number density used here is
$n= 2 \times 10^{13} {\rm cm}^{-3}$. (c) The transition point $\delta_c$ (black dotted)
and the corresponding energy $E_c$ (blue solid) vary with particle density, highlighting the lack of
universality of OFR. (d) The transition point $1/(k_F a_c)$ shows a non-monotonic dependence
on the effective range $k_F |r_0|$.}
\label{fig:transition}
\end{figure}
With the knowledge of molecule and polaron solutions, we show in Fig.~\ref{fig:transition}(a) the
energies for both the polaron and the molecule in the shallow branch,
from which one can see clearly that there exists a polaron--molecule transition at
$\delta_c/E_F \approx -2.3$. This translates to a scattering length $1/(k_F a_c) \approx 0.81$
for the parameters we have chosen [Fig.~\ref{fig:transition}(b)].
When $\delta < \delta_c$, the ground state of the system
is a polaron with zero momentum, as one would expect for a weakly interacting
Fermi gas in the BCS limit. On the other hand, for $\delta > \delta_c$, the molecular state
with a zero center-of-mass momentum is more favorable. Notice that the polaron--molecule transition point
$\delta_c$ is smaller than the zero-momentum polaron to finite-momentum polaron transition
point $\delta_p$, which indicates that the
finite-momentum polaron state, although could be energetically favorable against a zero-momentum
polaron state in some parameter regime, remains only metastable. We stress that by considering particle-hole
fluctuations in the molecular ansatz, the polaron--molecule transition would be shifted
toward the BCS side, leading to an even larger stability region of the molecule.

The variation of the polaron--molecule transition point with respect to the atomic number density is shown in
Fig.~\ref{fig:transition}(c), from which one can see that $\delta_c$ increases with the particle
density. This clearly shows that the system does not have a universal behavior around
the resonance point, as the Fermi energy in this system is comparable to the differential Zeeman
splitting $\delta$ and can alter the scattering process.

As the lack of universality implies a narrow resonance, we plot in Fig.~\ref{fig:transition}(d)
the polaron--molecule transition point as a function of the effective range $r_0$, which is
determined by~\cite{ren2}
\begin{eqnarray}
\label{eq:r0}
r_0 = - \frac{a_{s1}^2}
{\sqrt{m |\delta|/\hbar^2} \left[ a_{s0} - \sqrt{m |\delta|/\hbar^2} (a_{s0}^2 - a_{s1}^2) \right]^2}.
\end{eqnarray}
As compared to the previous results obtained for narrow Feshbach resonances~\cite{massignan-12, castin-12},
where the transition point $1/(k_F a_c)$ decreases monotonically with increasing $k_F |r_0|$, in the present case, we observe
a non-monotonic dependence around the resonance [see the inset of Fig.~\ref{fig:transition}(d)].
This behavior corresponds to the divergence of $|r_0|$ as $\delta \to 0$~\cite{ren2}, in which limit
the thresholds of the open and closed channels become degenerate. We stress that such a
feature is a key difference from the case of a magnetic Feshbach resonance, where the open and the
closed channels are differentiated by the electronic-spin states, with far-detuned thresholds around
the resonance point.

\begin{figure}[t]
\centering{}
\includegraphics[width=0.9\columnwidth]{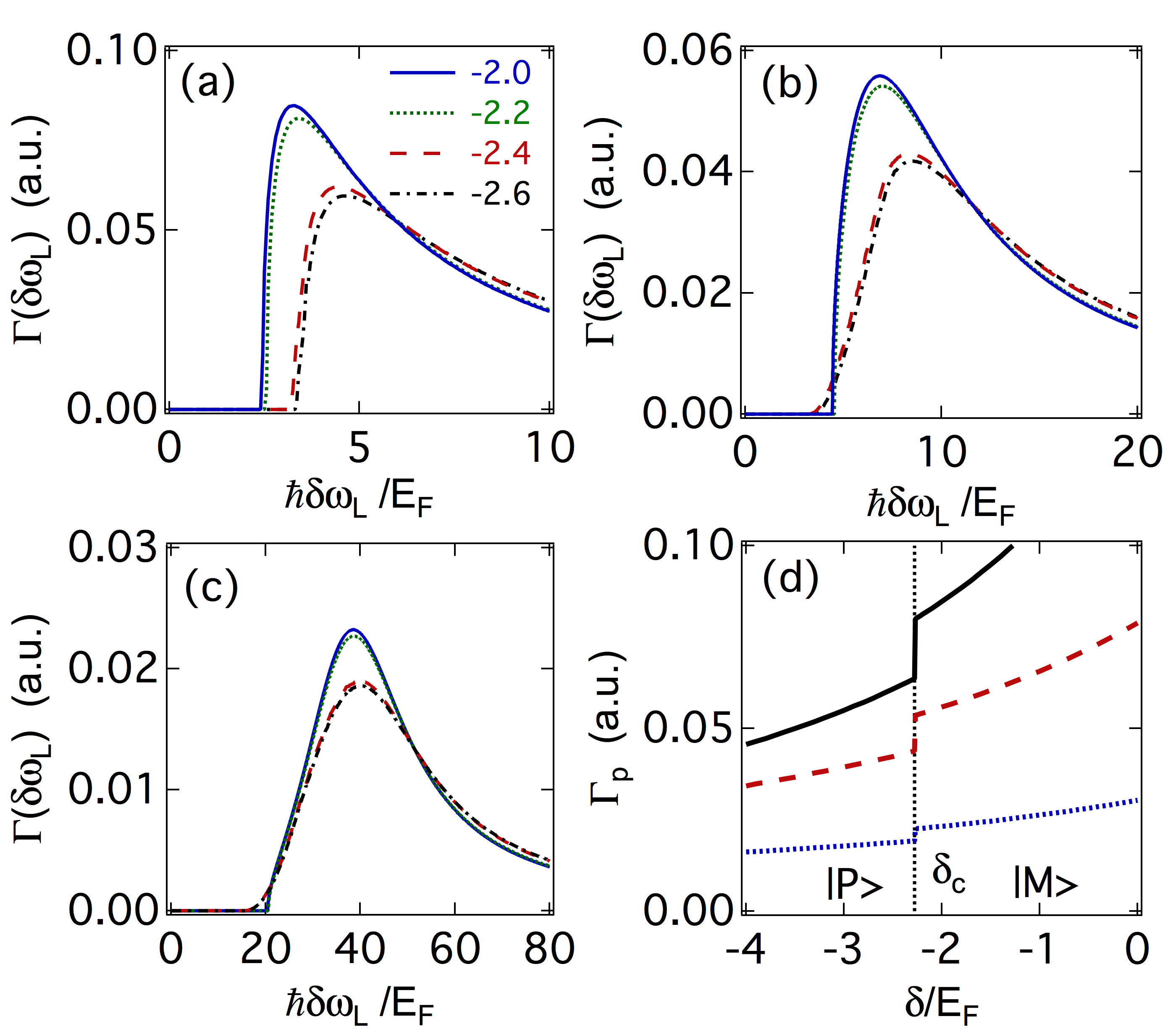}
\caption{(Color online) (a-c) Raman spectra with $\delta/E_F=-2.0$ (blue solid), $\delta/E_F=-2.2$ (green dotted), $\delta/E_F=-2.4$ (red dashed), and $\delta/E_F=-2.6$ (black dash-dotted). As the polaron-molecule transition point is at $\delta_c \approx -2.3$, the dashed and solid lines correspond to cases of polaron and molecule, respectively. The momentum recoil of the Raman process is (a) $k_R/k_F=0$, (b) $k_R/k_F=2$, and (c) $k_R/k_F=6$. (d) Peak value of the ground-state Raman spectra with changing $\delta$, with $k_R/k_F=0$ (black solid), $k_R/k_F=2$ (red dashed), and $k_R/k_F=6$ (blue dotted). Other parameters used are the same as in Fig.~\ref{fig:mol}.}
\label{fig:abs}
\end{figure}

In alkaline atoms, the polaron--molecule transition has been detected experimentally using radio-frequency (r.f.) spectroscopy~\cite{mitpolaron,grimmpolaron}. This is done by coupling the impurity state to a third by-stander state using an r.f. field~\cite{chinrfexp, chin05}. In $^{173}$Yb atoms, however, r.f. spectroscopy in which the impurity is coupled to another state in the same hyperfine manifold can be challenging, due to the particularly small Land\'e $g$-factor of the clock states. Instead, we propose to detect the polaron--molecule transition using Raman spectroscopy, where one of the states in $\{|g\downarrow\rangle, |g\uparrow\rangle, |e\downarrow\rangle, |e\uparrow\rangle\}$ is coupled to a by-stander state by a two-photon Raman process. For states in the $^1S_0$ manifold, a Raman process via the $^3P_1$ manifold can be used to transfer the population to another hyperfine state in the $^1S_0$ manifold. The transferred population can then be measured by coupling the state to the $^1P_1$ manifold. For states in the $^3P_0$ manifold, a Raman process via a highly excited $S$ or $D$ manifold can be used to transfer the population to the corresponding hyperfine state in the $^3P_1$ manifold. The transferred population can be measured by coupling the $^3P_1$ manifold to higher states and by monitoring the resulting fluorescence. As a concrete example, we discuss the former case, where $|g\uparrow\rangle$ is coupled to a third state by a Raman process.
The Hamiltonian accounting for this process can then be written as
\begin{eqnarray}
\label{eqn:abs}
V_{\rm ph} = V_0 \sum_{\bf k}
\left(a_{{\bf k}+k_R {\bf e}_x,3}^\dagger a_{{\bf k}g\uparrow} + {\rm H.C.} \right),
\end{eqnarray}
where $V_0$ is the effective Rabi frequency, $k_R {\cp e}_x$ is the momentum recoil of the Raman process, and H.C. stands for Hermitian conjugate.

According to the Fermi's golden rule, the total spectrum can be evaluated as
\begin{eqnarray}
\label{eqn:Fermigolden}
\Gamma(\delta\omega_{L}) = \sum_{f,i}\left| \langle \psi_f | V_{\rm ph} | \psi_i \rangle \right|^2
\delta\left(  \hbar \delta\omega_L-E_f + E_i \right),
\end{eqnarray}
where $\delta\omega_L$ is the two-photon detuning of the Raman process,
and $| \psi_f \rangle$ and $| \psi_i \rangle$ are the initial and final states
with the corresponding energies $E_f$ and $E_i$.
The total spectrum should satisfy the condition $\int \Gamma(\omega) d\omega=1$.
The resulting spectrum for the molecular state is then given by
\begin{align}
\Gamma(\delta\omega_L)=V_0^2\sum_{\cp k}|\beta_{\cp k}|^2\delta(\hbar\delta\omega_L+E_M-\epsilon_{\cp Q-\cp k}-\epsilon_{\cp k+k_R {\bf e}_x}),
\end{align}
while for the polaron the spectrum is given by
\begin{align}
\Gamma(\delta\omega_L)&=V_0^2\sum_{\cp k\cp q}|\beta_{\cp k\cp q}|^2\nonumber\\
&\hspace{-1cm}
\times\delta(\hbar\delta\omega_L+E_P-\epsilon_{\cp Q+\cp q-\cp k}-\epsilon_{\cp k+k_R {\bf e}_x}+\epsilon_{\cp q}+\frac{\delta}{2}).
\end{align}

In Figs.~\ref{fig:abs}(a-c), we show typical examples of Raman spectroscopy for both the polaron and the molecule
cases close to the polaron--molecular transition. Note that although the spectra look similar, those of the polarons are significantly lower in magnitude than those of the molecules. This is because the spectra intensity is closely related to the weight of wave functions in the closed channel, which acquires a finite jump by changing from the polaron to the molecular state. Therefore, as the magnetic filed is swept across the polaron--molecule transition, the spectra of the many-body ground state undergo a sudden change across the transition point [see Fig.~\ref{fig:abs}(d)]. This would allow us to identify the transition experimentally.

Alternatively, the polaron--molecule transition, as well as the polaron residue $\gamma$ of the polaron, can be probed by driving Rabi oscillations between the polaron and an initial non-interacting state~\cite{grimmpolaron,scazza-16}. The initial non-interacting state can be prepared by loading an impurity into a state $|3\rangle$ in the $^3P_1$ manifold above the Fermi sea $|g\downarrow\rangle_N$, where the impurity should have negligible interactions with states in $\{|g\uparrow\rangle, |g\downarrow\rangle, |e\uparrow\rangle, |g\downarrow\rangle\}$. One may then couple the non-interacting state and the polaron either using a r.f. field, or co-propagating Raman lasers. The coupling field can be written as 
\begin{eqnarray}
V_{c}=\Omega_0 \sum_{\cp k} (a^{\dag}_{\cp k,e\uparrow}a_{\cp k,3}+{\rm H.C.}).
\end{eqnarray}
Considering an initial state with zero-momentum impurity $|I\rangle=a^{\dag}_{q=0,3}|g\downarrow\rangle_N$, the effective Rabi frequency of the resulting coherent Rabi oscillation between $|I\rangle$ and the polaron $|P\rangle$ is $\Omega=\langle P|V_c|I\rangle=\gamma \Omega_0$. It is also straightforward to show that $\langle M|V_c|I\rangle=0$, such that there would be no Rabi oscillation should the system in a molecular state $|M\rangle$. Therefore, the coherent Rabi oscillation not only offers an unambiguous signal for the onset of the polaron--molecule transition, but also allows one to characterize the polaron residue $\gamma$.

\section{Conclusion}
\label{sec:con}
We study the impurity problem in a Fermi gas of $^{173}$Yb atoms near the orbital Feshbach resonance. Due to the spin-exchanging nature of the underlying two-body interactions, the formation of both the polaron and the molecule can have interesting spin-flipping processes. These lead to the occupation of originally empty hyperfine states. We show that a polaron to molecule transition exists close to the orbital Feshbach resonance, which can be detected experimentally using Raman spectroscopy. Our findings can be confirmed under current experimental conditions. Finally, we note that in the present work, we only focus on the attractive polarons. 
Repulsive branch of polarons, which has been investigated near magnetic Feshbach resonances~\cite{grimmpolaron, kohlpolaron, scazza-16}, may also be stabilized near an OFR, for which the spin-exchange interaction of an OFR may have interesting implications.

\acknowledgments
This work is supported by the NKBRP (2013CB922000), National Key R\&D Program (Grant No. 2016YFA0301700), the National Natural Science Foundation of China (Grant Nos. 60921091, 11274009, 11374283, 11434011, 11522436, 11522545), and the Research Funds of Renmin University of China (10XNL016, 16XNLQ03). W. Y. acknowledges support from the ``Strategic Priority Research Program(B)'' of the Chinese Academy of Sciences, Grant No. XDB01030200.


\end{document}